\documentclass[aps,prf,preprint,groupedaddress,longbibliography]{revtex4-1}

\usepackage{xcolor,bm}
\usepackage{bm,mathtools,esint}
\usepackage{graphicx,booktabs}
\usepackage{subfig,floatrow}
\usepackage{cleveref}
\newcommand{\red}[1]{\textcolor{black}{#1}}
\newcommand\Rey{Re}  
\newcommand\Pran{Pr}  
\newcommand\St{\mbox{St}}  
\newcommand\Nu{\mbox{Nu}}  
\newcommand\urms{u_{rms}} 
\newcommand\x{\bm{x}} 

\begin{document}


\title{Particle-to-fluid heat transfer in particle-laden turbulence}


\author{Hadi Pouransari}
\author{Ali Mani}
\email{alimani@stanford.edu}
\affiliation{Department of Mechanical Engineering, Stanford University,
Stanford, CA 94305, USA}


\begin{abstract}
Preferential concentration of inertial particles by turbulence is a well recognized phenomenon. This study investigates how this phenomenon impacts the mean heat transfer between the fluid phase and the particle phase. Using direct numerical \red{simulations of homogeneous and isotropic turbulent flows} coupled with Lagrangian point particle tracking, we explore this phenomenon over wide range of input parameters. Among the nine independent dimensionless numbers defining this problem, we show that particle Stokes number, defined based on large eddy time, and a new identified number called heat mixing parameter have the most significant effect on particle to gas heat transfer, while variation in other non-dimensional numbers can be ignored. \red{An investigation of regimes with significant particle mass loading, suggests that the mean heat transfer from particles to gas is hardly affected by momentum two-way coupling.} Using our numerical results we propose an algebraic reduced order model for heat transfer in particle-laden turbulence.
\end{abstract}


\maketitle

\section{Introduction}\label{sec:introduction}
A broad range of natural and industrial processes involve interaction of particles and background turbulent flows ---formation of clouds \citep{shaw2003particle,grabowski2013growth}, dispersion of pollutant in urban areas \citep{britter2003flow}, planetary accretion \citep{cuzzi2001size}, spray combustion \citep{sahu2014droplet}, and particle-based solar receivers \citep{pouransari2017effects} to name a few. 

Particle-turbulence interaction results in a range of well-studied phenomena. A particle immersed in turbulent flow experiences a centrifugal force from high vorticity regions toward high strain regions. This results in inhomogeneous distribution of particles, known as preferential concentration \citep{lazaro1989particle,squiresE91}. Where gravity is present particles exhibit preferential sweeping  \citep{wangM93}.
In wall-bounded particle-laden flows turbophoresis, which refers to the tendency of particles to concentrate close to the wall, is expected \citep{reeks1983transport}.

In many of the particle-laden flow scenarios, a primary interest is in understanding of thermal exchanges between the two phases. For example, in particle-based solar receivers, particles are the primary absorbers of external radiation, which then conductively transfer their absorbed heat to the carrier fluid. The heated particles absorb fraction of the received flux and transfer the rest to the surrounding fluid. \red{In this case radiation is not primarily absorbed by the gas phase since most gases are transparent to light}.

In the case of heated particle-laden flows, additional phenomena are observed. \cite{pouransari2017spectral} showed that under sufficiently large thermal flux, hot particles can modify turbulence spectra through pressure-dilatation. When gravity is present, heated particles give rise to non-uniform buoyant forcing of the flow, resulting in a sustained turbulence \citep{zamansky2014radiation,zamansky2016turbulent}. \cite{frankel2016settling} showed that when particles are heated the preferential sweeping can be supressed or even reversed. 

In a previous study \citep{pouransari2017effects} we investigated a specific regime of particle-laden flows and showed that preferential concentration of particles by turbulence can adversely impact the heat transfer efficiency. To obtain a fundamental understanding of impact of particle clustering on heat transfer, in the present study we consider a canonical setting involving heat transfer from inertial particles to \red{statistically stationary homogeneous isotropic turbulent flows}. By considering a combination of DNS data and a simple phenomenological model, we develop and verify an algebraic model for heat transfer in particle-laden turbulent flows. \red{Turbulence in this study is maintained by a forcing mechanism \citep{rosales2005linear}. We show that momentum two-way coupling between particles and fluid does not affect mean heat transfer between two phases.} 

\section{Model problem}\label{sec:modelProblem}
\subsection{Assumptions}\label{sec:assumptions}
We consider direct numerical simulations of homogenous isotropic turbulence (HIT) laden with heated point particles in a triply periodic box with length $L$. The simulation code \citep{pouransari2015parallel} is fourth order in time and second order in space using uniform staggered grid. A linear forcing scheme \cite{rosales2005linear,lundgren2003annual} is used to maintain a statistically stationary turbulence with zero-mean velocity.

\red{Each simulation starts with two transition stages. Collection of heat transfer statistics is performed after these transitions when thermal exchange process reaches a statistically stationary condition. At first transition stage} the cold mixture (with temperature $T_0$) is simulated for sufficiently large time with no external heating to obtain a fully developed particle-laden turbulence. By monitoring the fluid kinetic energy and particle segregation \citep[see for example][]{vie2016particle} versus time, we verify \red{a healthy particle-laden turbulence is achieved}. This is achieved after 50 large-eddy turnover times defined below. \red{The first stage is followed by the second transition stage, where} particle heating is activated with constant heat for each particle, and the heated mixture is allowed to be developed. The statistically stationary heated state is verified by monitoring the mean particle to fluid heat flux versus time. All statistics are collected after these two transition stages over a period of order 100 large-eddy turnover times. Note that when statistically stationary state is achieved, the ensemble-averaged particle and fluid temperatures grow linearly with time, while the  ensemble-averaged temperature difference and particle-to-fluid heat transfer are constant.

The fluid phase is assumed to be variable density governed by the ideal gas equation of state $P = \rho R T_g$, subject to a low-Mach flow. \red{Therefore, the thermodynamic pressure is considered to be constant in space while can change in time.} The dynamic viscosity $\mu$, constant-volume and constant-pressure specific heat coefficients $C_v$, $C_p$, and heat conductivity coefficient $k$ of the gas are assumed to be constant and independent of the temperature. Conservation of mass implies constant averaged gas density $\rho_0$.

Mono-dispersed spherical particles with density $\rho_p \gg \rho_0$ and a constant diameter $d_p$ much smaller than the Kolmogorov micro scale are suspended in the fluid. Particle specific heat coefficient ${C_v}_p$ is constant and independent of the temperature. We assume particle temperature $T_p$ to be a lumped quantity (constant along one particle) justified by large particle Biot number. \red{The slip velocity between particle and the surrounding gas is assumed to be a finite small value. Thus, we assume that the particle momentum and heat exchange with the fluid can be expressed, respectively, in terms of drag and heat exchange laws derived in the low Reynolds and P\'eclet limits. We ignore the convective effects at the scale of particles justified by low thermal P\'eclet number based on slip velocity, and particle diameter.}

The particle-fluid mixture is assumed to be very dilute (volume fraction $\sim10^{-5}$). We use a simplified version of the \citet{maxey1983equation} equations describing the dynamics of an immersed particle, and model particle motion through the Lagrangian point particle framework. In the regimes considered in our study, Stokes drag is the only significant force experienced by particles. We ignore momentum two-way coupling (i.e., particles do not modify the fluid through momentum equation). \red{In Section \ref{sec:TWC} we will revisit the impact of momentum two-way coupling on mean heat transfer.}

\subsection{Non-dimensional equations}\label{sec:nondimeqn}
Based on the aforementioned assumptions, we introduce a set of dimensionless equations describing heated particle-laden flows. We use the flow integral length scale $l = \urms^3 / \epsilon$ as the reference length scale, where $\urms$ is the root mean square of the single-component velocity fluctuations, and $\epsilon$ is the averaged dissipation rate. Large-eddy turnover time $\tau_l = l / \urms$ is used as the reference timescale. $\urms$, $\rho_0$, and $T_0$ are used to non-dimensionalize velocities, gas density, and temperatures, respectively.

The conservation of mass, momentum, and energy for the gas is represented by set of non-dimensional \cref{eqn:GCOM,eqn:GCOMM,eqn:GCOE}, respectively.
\begin{equation} \label{eqn:GCOM}
\frac{\partial \rho}{\partial t}+\frac{\partial}{\partial x_j}(\rho u_j)=0
\end{equation}
\begin{equation} \label{eqn:GCOMM}
\frac{\partial} {\partial t}(\rho u_i)+\frac{\partial}{\partial x_j}(\rho {u}_i {u}_j)=-\frac{\partial p}{\partial x_i} + \frac{1}{\Rey} \frac{\partial}{\partial x_j}\left(\frac{\partial {u}_i}{\partial x_j}+\frac{\partial {u}_j}{\partial x_i}-\frac{2}{3}\frac{\partial {u}_k}{\partial x_k}\delta_{ij}\right) + \frac{1}{3} \rho u_i
\end{equation}
\begin{equation} \label{eqn:GCOE}
\frac{\partial}{\partial t} (\rho T_g)+\gamma \frac{\partial}{\partial x_j}(\rho T_g {u}_j)= \frac{\gamma}{\Rey \Pr} \frac{\partial^2 T_g}{\partial x_j \partial x_j} + \sum_{i=1}^{N_p} \frac{T_{p_i} - T_g(x_{p_i})}{\sigma_l} \delta(\x - \x_{p_i}) 
\end{equation}
In the above equations $\rho$, $u$, $p$, and $T_g$ denote density, velocity, hydrodynamic pressure, and temperature of the gas phase, respectively. The last term in \cref{eqn:GCOMM} is the non-dimensionalized linear forcing term \citep{rosales2005linear} to maintain turbulence (with dimensionless dissipation rate 1). The last term in \cref{eqn:GCOE} is the heat transfer from particles to gas. $T_{p_i}$ and $x_{p_i}$, respectively, represent temperature and position of a particle with index $1 \le i \le N_p$, where $N_p$ is the total number of Lagrangian particles. $\delta$ is dimensionless three-dimensional Dirac delta function, \red{which is numerically approximated using trilinear interpolation and projection.}

The non-dimensionalized equation of state for ideal gas is $\rho T_g =P$. The thermodynamic pressure is non-dimnesionalized with $\rho_0 R T_0$, where $R$ is the gas constant. Note that given the low Mach number assumption, the thermodynamic pressure is assumed to be constant in space.

\Cref{eqn:PCOMM,eqn:PCOE} are set of non-dimensional equations representing kinematics, dynamics, and energy conservation for a particle $p_i$, respectively.
\begin{equation} \label{eqn:PCOMM}
\frac{d}{dt} \x_{p_i} = v_{p_i}, \quad  \frac{d}{dt} v_{p_i} = - \frac{ {v}_{p_i} - {u}(\x_{p_i})}{\St_l}
\end{equation}
\begin{equation} \label{eqn:PCOE}
\chi \frac{d}{dt} T_{p_i} = \mathcal{S} - \frac{T_{p_i} - T_g(\x_{p_i})}{\sigma_l}
\end{equation}
In the above equations $v_{p_i}$ denotes velocity of a particle with index $i$. The first and second terms in the right hand side of \cref{eqn:PCOE} are the constant heat flux absorbed by a particle and the heat transfer from particle to gas, respectively.

Next, we introduce the dimensionless factors denoted in \cref{eqn:GCOM,eqn:GCOMM,eqn:GCOE,eqn:PCOMM,eqn:PCOE}. $\Rey = \rho_0 \urms l/\mu$ is the Reynolds number. $\gamma = C_p / C_v$ is the ratio of gas heat capacities. $\Pran = C_p \mu / k$ is the Prandtl number. $\St_l = \tau_p / \tau_l$ is the particle Stokes number, which is the ratio of particle momentum relaxation time $\tau_p = \rho_p d_p^2 / (18 \mu)$ to the gas large-eddy turnover time.

$\sigma_l = \tau_{th} / \tau_l$ is the heat mixing parameter defined as the ratio of gas thermal relaxation time $\tau_{th} = \rho_0 C_v/(\Nu \pi d_p k n_0)$ to the large-eddy turnover time, where $n_0 = N_p/L^3$ is the mean particle concentration. $\Nu$ is the Nusselt number for particle to gas heat transfer. Note that we selected the fluid thermal relaxation time, as opposed to the commonly used particle thermal relaxation time to form the dimensionless heat mixing parameter. For reasons that are discussed in \cref{sec:closure} we shall see the former choice results in minimal number of significant dimensionless parameters, while the latter choice would not.

$\chi = n_0 {C_v}_p \rho_p \pi d_p^3/(6 \rho_0 C_v)$ is the ratio of dispersed phase total heat capacity to the gas phase total heat capacity. $\mathcal{S} = \mathcal{H} n_0 \tau_l / (T_0 \rho_0 C_v)$ is the non-dimensional heat source, where $\mathcal{H}$ is the external heat flux received by each particle. $\mathcal{S}$ can be interpreted as the ratio of large-eddy turnover time to the gas warmup time $\tau_h = T_0 \rho_0 C_v / (\mathcal{H} n_0)$.

\section{Heat-transfer model}\label{sec:averaged}
\subsection{Reduced order equations}\label{sec:averaging}
In this section, we develop a reduced order model to describe the evolution of the averaged particle and gas temperature. We start with definition of the averaging operator.

For a given scalar fields $\psi$ and a weight function $w$ in domain $\Omega$, the weighted average of $\psi$ is defined as
\begin{equation}\label{eqn:LA}
\langle \psi \rangle_w = \frac{\langle w \psi \rangle}{\langle w \rangle}, \text{ where } \langle \psi \rangle = \frac{1}{\text{Vol} (\Omega)} \iiint_{\Omega} \psi(\x) d\x.
\end{equation}
If we use gas density $\rho$ as weight, $\langle \psi \rangle_{\rho}$ is the Favre-average \citep{wilcox1998turbulence}. We define dimensionless particle local concentration as $n(\x) = 1/N_p \sum_{i=1}^{N_p} \delta ( \x - \x_{p_i})$. Therefore, $\langle \psi \rangle_n = 1/N_p \sum_{i=1}^{N_p} \psi(\x_{p_i})$. In case of particle temperature $ \langle T_p \rangle_n = 1/N_p \sum_{i=1}^{N_p} T_{p_i}$.

To obtain reduced order heat transfer equations for gas and particles, we take average of \cref{eqn:GCOE,eqn:PCOE}. Noting that $\langle \rho \rangle$ = $\langle n \rangle = 1$ in the non-dimensional form we get
\begin{equation}\label{eqn:1DGEP}
\begin{aligned}
\frac{d}{dt} \langle T_g \rangle_{\rho} &= \frac{\langle T_p \rangle_n - \langle T_g \rangle_n}{\sigma_l}\\
\chi \frac{d}{dt} \langle T_p \rangle_n &= \mathcal{S} -  \frac{\langle T_p \rangle_n - \langle T_g \rangle_n}{\sigma_l}
\end{aligned}
\end{equation}
\subsection{Correction factor}
\Cref{eqn:1DGEP} is exact but not closed due to appearance of $\langle T_g \rangle_n$ on the right hand sides. Similar to \cite{pouransari2017effects}, we define a correction factor for the heat transfer term as follows to close the equations.

\begin{equation} \label{eqn:cf}
\varphi = \frac{\langle T_p \rangle_n - \langle T_g \rangle_n}{\langle T_p \rangle_n - \langle T_g \rangle_{\rho}}
\end{equation}
Therefore, \cref{eqn:1DGEP} transforms to the following equations.
\begin{equation}\label{eqn:1DGPEphi}
\begin{aligned}
\frac{d}{dt} \langle T_g \rangle_{\rho} &= \varphi \frac{\langle T_p \rangle_n - \langle T_g \rangle_{\rho}}{\sigma_l}\\
\chi \frac{d}{dt} \langle T_p \rangle_n &= \mathcal{S} -  \varphi \frac{\langle T_p \rangle_n - \langle T_g \rangle_{\rho}}{\sigma_l}
\end{aligned}
\end{equation}
$\langle T_g \rangle_n$ is the average gas temperature seen by particles, similar to the concept of average gas velocity seen by particles when the drift velocity is concerned \citep{jung2008behavior}. Since particles are directly heated, it is expected that  the average temperature of the gas at the location of particles to be greater than the volume-averaged temperature of the gas, $\langle T_g \rangle_n \ge \langle T_g \rangle_{\rho}$. Hence, we expect $0 \le \varphi \le 1$. The closure question is then to determine $\varphi$ in terms of known input parameters. This is similar to the work by \citet{sumbekova2017preferential}, in which they investigated the parameter space of unheated particle-laden turbulence experimentally to explore the effect of each parameter on preferential clustering.

\subsection{Parameter study}\label{sec:paramStudy}
We investigate the dependence of the parameter $\varphi$ in \cref{eqn:1DGPEphi} on the dimensionless numbers governing the problem as introduced in \cref{sec:nondimeqn}. Considering common gas-solid mixtures, we assume $\gamma = 1.4$, $\Pran = 0.7$, and $\Nu = 2$. We sweep the parameter space by changing the remaining dimensionless number(s) of interest at a moment, while all other dimensionless numbers are kept constant. In \cref{tab:nominal} we list the nominal value and sweeping range of each dimensionless number. For each dimensionless number, we run a simulation of the full 3D equations for sufficiently long time, and average the value of $\varphi$ when the heated HIT is developed in time by post-processing the data to compute terms on the right-hand-side of \cref{eqn:cf}.

\red{Note that we run very long simulations as a mathematical trick to compute converged statistics more easily instead of computing many simulations over short time and then taking their average. For this we cosider constant material properties. However, our data is intended to represent the much shorter evolution of ensemble averaged statistics in an experiment.}

\begin{table}[hbt!]
  \begin{center}
\def~{\hphantom{0}}
  \begin{tabular}{lcccccc}
  {\bf parameter:}&$N_p$&$\mathcal{S}$&$\chi$&$\St_l$&$\sigma_l$&$\Rey$\\\midrule[2pt]
  {\bf nominal:} &$10^5$ & 2 & 1.0 & 0.15 & 0.5 & 47\\
  {\bf range:} &$[10^5~10^7]$ & $[1 ~ 10]$ & $[10^{-2} ~  10^2]$& $[10^{-2} ~ 1]$& $[10^{-2}~10^2]$ & $[10~10^3]$\\
  \end{tabular}
  \caption{Nominal value of dimensionless numbers.}
  \label{tab:nominal}
  \end{center}
\end{table}

\begin{figure}[hbt!]
\centerline{\includegraphics[width=.8\textwidth]{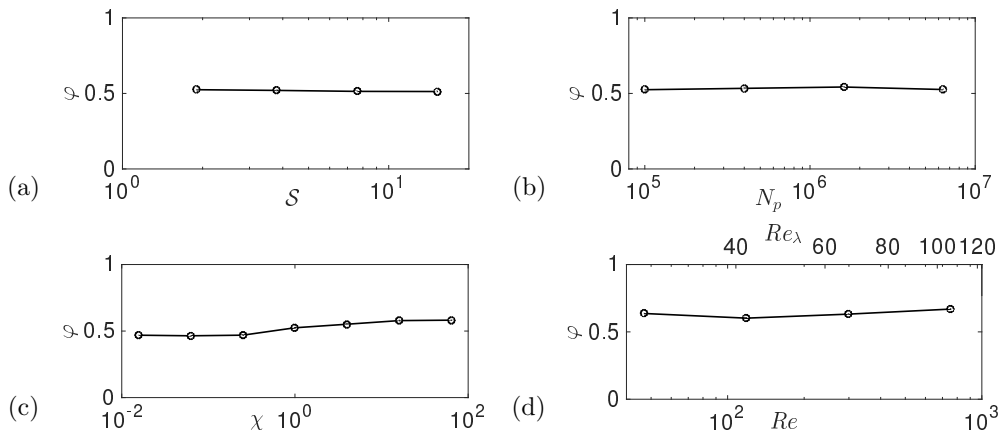}}
\caption{\label{fig:notimp} $\varphi$ as a function of (a) total number of particles $N_p$; (b) Non-dimensional heat flux $\mathcal{S}$; (c) particle to gas heat capacity ratio $\chi$; (d) Reynolds number $\Rey$ and $\Rey_{\lambda}$.}
\label{fig:notimp:S}
\label{fig:notimp:Np}
\label{fig:notimp:chi}
\label{fig:notimp:Re}
\end{figure}

The heat transfer equations are linearly dependent on $\mathcal{S}$, and thus one expects that the correction factor, $\varphi$, to be independent of this parameter. This expectation is justifiable as long as thermal flux is not too strong to modify the turbulence itself \citep{pouransari2017spectral}.
\Cref{fig:notimp:S}a depicts $\varphi$ as a function of $\mathcal{S}$, while other dimensionless numbers are kept at their nominal values, verifying $\varphi$ is independent of $\mathcal{S}$. In addition, in the limit of sufficiently large $N_p$ and negligible particle-particle collisions, it is expected that $\varphi$ to be independent of $N_p$. \Cref{fig:notimp:Np}b shows $\varphi$ versus $N_p$ confirming its independence on $N_p$. Increasing $N_p$ and/or $\mathcal{S}$ gives rise to higher total heat transfer from particles to gas, yet the correction factor is constant.

In \cref{fig:notimp:chi}c we plot $\varphi$ as a function of particle to gas total heat capacity $\chi$. This figure suggests weak dependence of $\varphi$ on $\chi$ particularly in the limit of large or small $\chi$. In \cref{sec:closure} we provide a phenomenological model justifying this observation. Therefore, only three remaining dimensionless numbers $\Rey, \St_l$, and $\sigma_l$ may significantly affect the correction coefficient $\varphi$.

\Cref{fig:notimp:Re}d illustrates variation of $\varphi$ as a function of Reynolds number. The bottom x-axis shows $\Rey$, and the top x-axis shows $\Rey_{\lambda} = \rho_0 \urms \lambda / \mu$, the Reynolds number based on the Taylor's micro-scale. For HIT we have $\Rey = {\Rey_{\lambda}^2}/{15}$. The value of other dimensionless numbers are the same as \cref{tab:nominal}, except the particle Stokes number which is kept at $\St_l = 0.06$. Our results suggest that when Stokes number and heat mixing parameter are defined based on the large eddy turnover time, $\varphi$ is a weak function of the Reynolds number. We further discuss implications of alternative choices of reference time scale in the definition of Stokes number and heat mixing parameter in \cref{sec:timescale}.

Note that we found the correction factor $\varphi$ to be independent of $N_p$, $\mathcal{S}$, and $\chi$ assuming no turbulence modification by the particles. In general, particles can modify the background turbulence either through the momentum exchange or through local expansion resulting from heat transfer. The former is significant when particle mass loading ratio is high \citep{squires1990particle, elghobashi1993two, sundaram1999numerical}, and the latter is significant in case of high heating that results in high dilatational modes quantified by $|\tau_l \nabla \cdot u|$ \citep{pouransari2017spectral}. This is the case in most turbulent combustion applications, for example. \red{All of the investigated cases are indeed in the regime where the dilatation due to heating is small compared to large eddy turnover time. However, the mass loading ratio is significant in some cases. However, we show in Section \ref{sec:TWC} that although turbulence modulation by particles can be considerable in these cases, impact on the mean heat transfer is negligible.}

\begin{figure}[hbt!]
\centerline{\includegraphics[width=.8\textwidth]{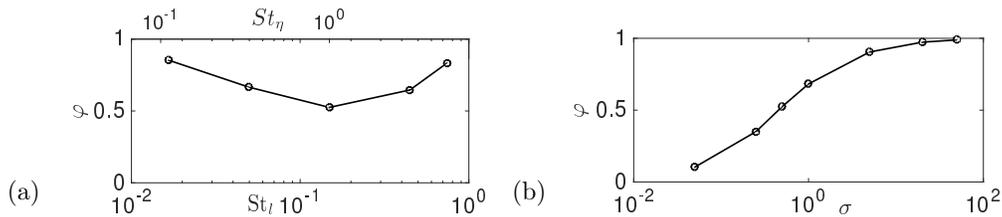}}
\caption{\label{fig:imp} $\varphi$ as a function of (a) Stokes number $\St_l$ and $\St_{\eta}$; (b) heat mixing parameter $\sigma_l$.}
\label{fig:imp:St}
\label{fig:imp:sigma}
\end{figure}

In \cref{fig:imp:St}a the heat transfer correction factor is plotted as a function of particle Stokes number. The bottom x-axis is Stokes number based on large-eddy turnover time $\St_l$, and the top x-axis shows the Stokes number based on the Kolmogorov time scale $\St_{\eta} = \tau_p / \tau_{\eta}$. For HIT we have $\St_{\eta} = \sqrt{\Rey}~\! \St_l$.

For very small and large Stokes numbers particle distribution is close to homogeneous. Therefore, the weights in \cref{eqn:LA} are almost uniform, and $\varphi$ is close to one. For moderate values of Stokes number the highest level of preferential concentration is expected. In this case particle to gas heat transfer occurs at the location of particle clusters that inevitably introduce spatial heterogeneity. Hence, when preferential concentration is high, the effective volume of cold gas seen by particles is reduced resulting in less heat transfer from particles to gas (i.e., $\varphi < 1$).

\Cref{fig:imp:sigma}b shows $\varphi$ as a function of heat mixing parameter $\sigma_l$. $\varphi$ is a strictly increasing function of $\sigma_l$, such that as $\sigma_l \to \infty$, $\varphi \to 1$, and as $\sigma_l \to 0$, $\varphi \to 0$. The value of $\sigma_l$ quantifies the rate of heat mixing by turbulence in terms of gas thermal relaxation time scale. Small values of $\sigma_l$ means heat mixing due to turbulence is weak, and large values of $\sigma_l$ represents strong heat mixing due to turbulence. We study the effects of simultaneous variations of $\St_l$ and $\sigma_l$ in \cref{sec:closure}.

\subsection{The choice of reference timescale}\label{sec:timescale}
In this study we use large-eddy turnover time as the flow reference timescale. Alternatively, we could use the Kolmogorov time scale as the reference flow timescale. Using the Kolmogorov time scale, in particular, is appealing in order to define the Stokes number as done in numerous previous studies characterizing preferential concentration.

In \cref{fig:notimp:Re}d we showed that the correction factor $\varphi$ has small dependence on Reynolds number, when all other non-dimensional numbers are kept constant. Note that as the Reynolds number increases the ratio between large-eddy turnover time and the Kolmogorov time also increases as well. Therefore, the choice of reference time scale for Stokes number and the heat mixing parameter, that are kept constant while Reynolds is varied, is important.

\begin{figure}[hbt!]
\centerline{\includegraphics[width=.95\textwidth]{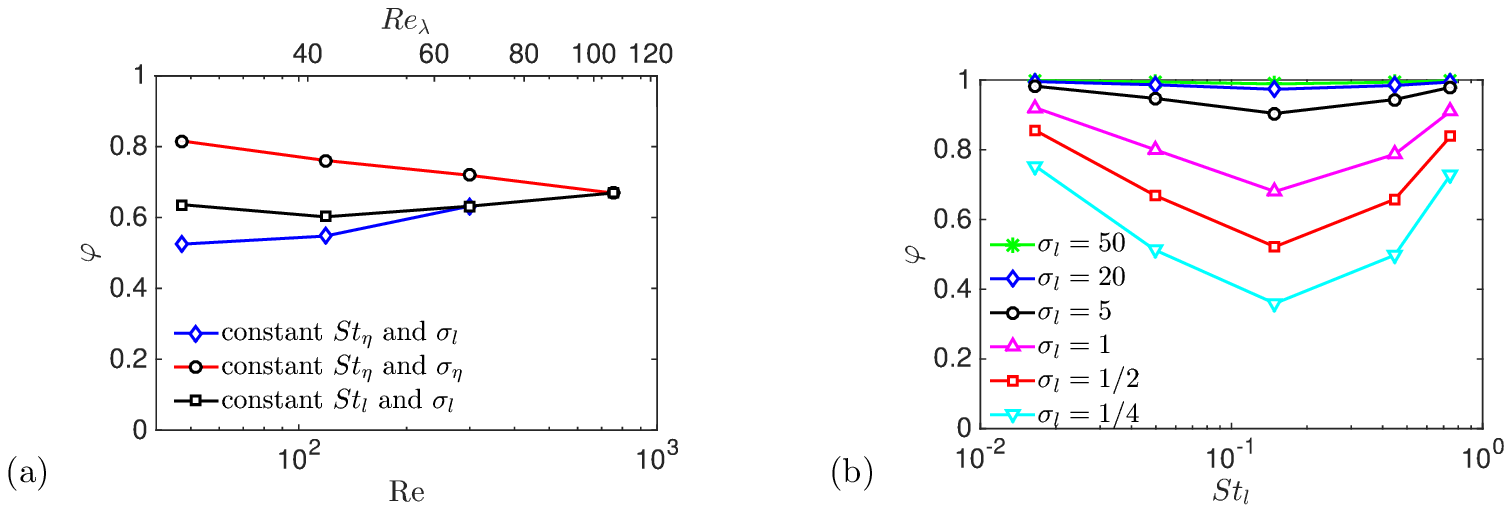}}
\caption{Variation of $\varphi$ as a function of (a) Reynolds number when Stokes number and heat mixing parameter are kept constant. \red{Subscripts $\eta$ and $l$ refer to normalizing by Kolmogorov and large-eddy turnover times, respectively: $St_l = \tau_p/\tau_l, St_{\eta} = \tau_p / \tau_{\eta}, \sigma_l = \tau_{th}/\tau_l, \sigma_{\eta}=\tau_{th}/\tau_{\eta}$;} (b) heat mixing parameter $\sigma_l$ and Stokes number $\St_l$.}
\label{fig:Re_study}
\label{fig:st_sigma}
\end{figure}

In \cref{fig:Re_study}a variation of $\varphi$ with Reynolds number is shown when different flow timescales are used for defining the Stokes number and/or heat mixing parameter $\sigma$. Subscript $l$ denote normalizing with the large-eddy turnover time, and subscript $\eta$ denote normalizing with the Kolmogorov time. \Cref{fig:Re_study}a demonstrates that our choice of large-eddy turnover time for both Stokes number and heat mixing parameter results in the least dependency of $\varphi$ on Reynolds. Further studies are required to investigate variations of $\varphi$ for much larger values of Reynolds number.

\subsection{Effect of momentum two-way coupling}\label{sec:TWC}
\red{In all cases considered above momentum two-way coupling between particles and gas is ignored in the simulations to simplify the analysis of the system. However, for significant mass loading ratios momentum exchange between the two phases can modify the turbulence dynamic.}

\red{To verify the validity of the results presented here, we ran a simulation of the nominal case (see \cref{tab:nominal}) with consideration of momentum two-way coupling between particles and gas \citep{elghobashi1993two}. Similar to heat exchange between two phases we use trilinear interpolation and projection for numerical calculation of momentum exchange. The correction factor, $\varphi$, for the nominal case changes from 0.525 to 0.526 when momentum two-way coupling is considered. This shows that the effect of momentum two-way coupling is negligible on the correction factor introduced here.}

\subsection{Closure model}\label{sec:closure}
We concluded that Stokes number and heat mixing parameter, defined based on $\tau_l$, are the most relevant dimensionless numbers determining $\varphi$. Therefore, we sweep the parameter space in two dimensions ($\St_l$ and $\sigma_l$) to discover the full dependency of $\varphi$ on the input parameters.  In \cref{fig:st_sigma}b variation of $\varphi$ as a function of Stokes number is plotted for different values of $\sigma_l$. Other dimensionless numbers are kept at their nominal values as reported in \cref{tab:nominal}. 

\begin{figure}[hbt!]
  \centerline{\includegraphics[width=.6\textwidth]{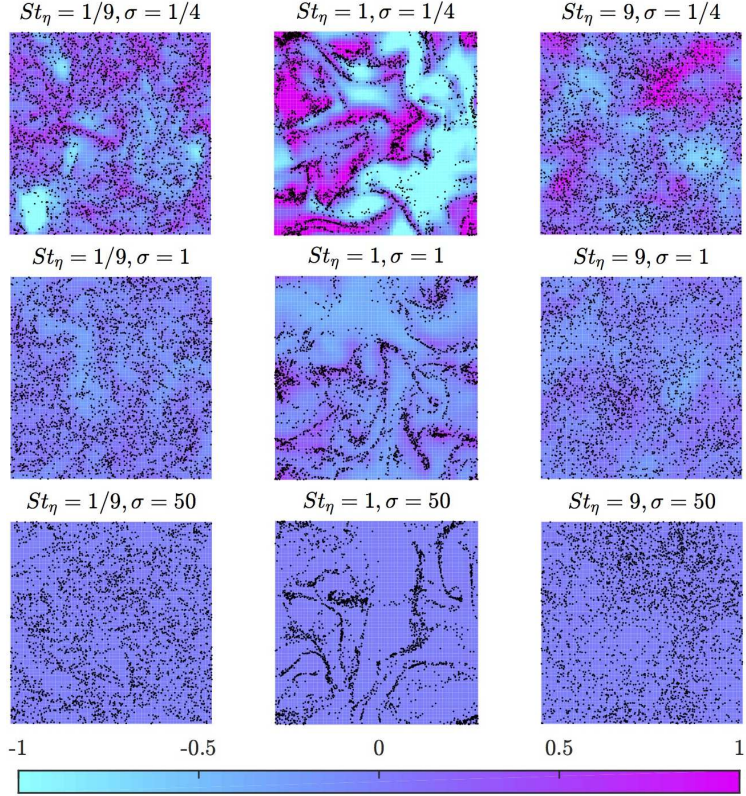}}
  \caption{Particle distribution and normalized fluid temperature deviations, $T_g - \langle T_g \rangle_{\rho}$, contours for different combinations of $\St_{\eta}$ and $\sigma_l$ in one slice of the domain.}
\label{fig:grid}
\end{figure}

The non-monotonic dependence of $\varphi$ as a function of $\St_l$ can be observed for all values of $\sigma_l$. However, as $\sigma_l \to \infty$ dependence of $\varphi$ on particle Stokes number vanishes. For large values of $\sigma_l$ the heat mixing due to turbulence is strong, therefore, even for high level of preferential concentration the heat transferred from particles to gas is quickly mixed uniformly. This fast mixing makes the gas temperature uniform despite heterogeneity of the source, thus brings $\langle T_g \rangle_n$ closer to $\langle T_g \rangle_{\rho}$ leading to $\varphi \simeq 1$. This effect is visually evident in \cref{fig:grid} where particle distribution and normalized fluid temperature deviations, $T_g - \langle T_g \rangle_{\rho}$, is illustrated for various combinations of $\St_{\eta}$ and $\sigma_l$. Note that particles have high level of preferential concentration when $\St_{\eta} \sim \mathcal{O}(1)$ irrespective of value of $\sigma_l$ (the middle column in \cref{fig:grid}). However, only for small values of $\sigma_l$ the particle preferential concentration affect particle to gas heat transfer.

Here, we introduce a simple phenomenological heat-transfer model consistent with the observed dependency of $\varphi$ on the dimensionless parameters. We use the result of this approach to provide a closed algebraic form for the correction coefficient $\varphi$.

\begin{figure}[hbt!]
  \centerline{\includegraphics[width=.99\textwidth]{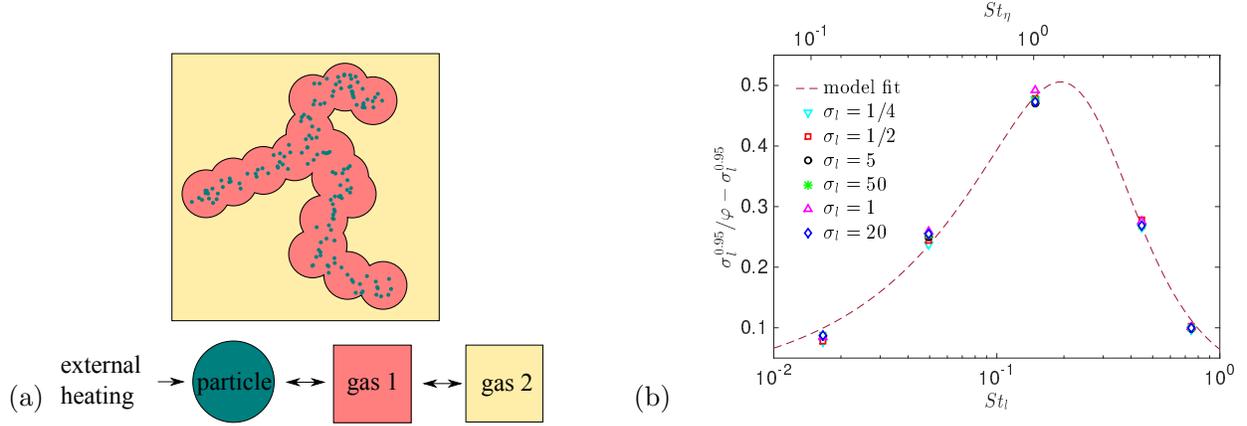}}
\caption{(a) Particle-to-gas heat transfer model schematics; (b) collapse of ${\sigma_l}^p /\varphi - {\sigma_l}^p$ versus $\St_l$ for different values of $\sigma_l$, with $p=0.95$.}
\label{fig:htm}
\label{fig:collapse}
\end{figure}

Consider the cloud of gas in vicinity of particles as shown in \cref{fig:htm}a. Assume this cloud occupies volume fraction $f$ of the total gas. The coefficient $0 \le f \le 1$ is only a function of particle spatial distribution and thus $f=f(\St_l)$. In this approach we consider one averaged temperature for cloud of gas near particles \red{(gas 1)}, and one for the rest of the gas \red{(gas 2)}, denoted by $T_{g_1}$ and  $T_{g_2}$, respectively. We assume particles receive energy from an external heat source, and transfer energy conductively to the surrounding cloud of gas. This cloud then transfers heat convectively to the rest of the gas. In the former, heat-transfer is dominated by gas thermal relaxation time $\tau_{th}$ augmented by factor $f$ to account for smaller mass fraction. In the latter, large-eddy turnover time $\tau_l$ is the dominant time scale in heat transfer as the mixing by large eddies is the main heat-transfer mechanism \citep[see][]{prandtl1925bericht,dimotakis2005turbulent,vassilicos2015dissipation}.  Using the same non-dimensionalization as in \cref{sec:nondimeqn} the governing equations are as follows.
\begin{equation}\label{eqn:model}
\begin{aligned}
\chi  \frac{d}{dt} \langle T_p \rangle_n &= -\frac{1}{\sigma_l}(\langle T_p \rangle_n - T_{g_1}) + \mathcal{S}\\
f \frac{d}{dt} T_{g_1} &= \frac{1}{\sigma_l}(\langle T_p \rangle_n - T_{g_1}) - (T_{g_1} - T_{g_2})\\
(1-f) \frac{d}{dt} T_{g_2} &= (T_{g_1} - T_{g_2})
\end{aligned}
\end{equation}
Neglecting spatial variation of gas density we can write:
\begin{equation}\label{eqn:model:Tg}
\langle T_g \rangle_{\rho} = f~\!T_{g_1} + (1-f)~\! T_{g_2}
\end{equation}
Combining \cref{eqn:model,eqn:model:Tg}, and noting that under the fully developed conditions all temperatures increase linearly with time with the same slope ($=\mathcal{S}/(\chi+1)$, to satisfy conservation of energy for the full system) we can derive the following equation:
\begin{equation}
\chi \frac{d}{dt} \langle T_p \rangle_n = \mathcal{S} - \frac{\sigma_l}{\sigma_l + (1-f)^2} \cdot \frac{\langle T_p \rangle_n - \langle T_g \rangle_{\rho}}{\sigma_l}
\end{equation}
Comparing against equation \cref{eqn:PCOE} suggests the correction factor $\varphi$ introduced in \cref{eqn:cf} is $\varphi = \sigma_l/(\sigma_l + (1-f)^2)$. This analysis 
also provides insights on the observed weak dependence of $\varphi$ on $\chi$, as the derived expression does not involve parameter $\chi$ in this simplified limit.

Next, assuming $f$ being only function of $\St_l$, we consider a generalized form for $\varphi = {\sigma_l}^p/({\sigma_l}^p +g(\St_l))$. Based on our phenomenological model we expect a value of $p$ close to 1. Reverse engineering of this expression indicates that ${\sigma_l}^p /\varphi - {\sigma_l}^p$ versus $\St_l$ must result in plots independent of $\sigma_l$. As shown in \cref{fig:collapse}b, $p=0.95$ results in collapse of curves corresponding to different values of $\sigma_l$ confirming the usefulness of our phenomenological model.

We consider $ g(\St_l) = \alpha~\! \St_l^{\eta_1} / (\St_l^{\eta_2} + \beta)$ as a  generalization to the expression suggested by \citet{esmaily2016analysis} to quantify the level of particle preferential concentration. We found the optimal values of $\alpha = 0.066$, $\beta = 0.025$, $\eta_1 = 0.8$, and $\eta_2 = 2.8$ by fitting $\varphi$ to our data \red{(see dashed line in \cref{fig:collapse}b)}. Finally, we propose the following algebraic form for particle-to-gas heat transfer correction coefficient:
\begin{equation} \label{eqn:closure}
\varphi = \frac{{\sigma_l}^{0.95}}{{\sigma_l}^{0.95} + 0.066~\! \frac{\St_l^{0.8}} {\St_l^{2.8} + 0.025}}
\end{equation}

\section{Conclusions}
In this paper, we developed a non-dimensional set of equations describing heated point particles suspended in a variable-density turbulent flow. We studied the averaged particle to fluid heat transfer using direct numerical simulations for fluid and Lagrangian point particle tracking for the dispersed phase. In the presented formulation, nine non-dimensional numbers appear. Considering wide range of applications involving small solid particles in gas, we investigated the regime in which $\gamma = 1.4$, $\Pran = 0.7$, and $\Nu = 2$, while we systematically varied all other non-dimensional numbers.

We showed that in \red{the regimes where the dilatational time is not fast compared to eddy time} the number of particles, dimensionless heating rate, ratio of particle to fluid heat capacities, and Reynolds number
have minor effect on mean heat transfer coefficient. However, our results indicate that particle Stokes number ($\St_{\eta}$) and the newly introduced non-dimensional number, the heat mixing parameter ($\sigma_l$), have significant effect on particle to fluid heat transfer. The former parameter governs the topology of particle distribution, which controls the spatial distribution of heat sources to the fluid. The latter parameter determines how fast the background fluid can mix the heat received by particles \red{compared to the gas nominal thermal relaxation time.} Therefore, even for high level of preferential concentration we can expect almost uniform fluid temperature distribution due to turbulent mixing when $\sigma_l$ is high.

Using our parameter study, in \cref{sec:closure} we introduced an algebraic closure formula to model macroscopic particle to gas heat transfer for general conditions. The inputs of our model are $\St_l$ and $\sigma_l$ that found to be the most relevant non-dimensional numbers for particle to fluid heat transfer. \red{This result can be used as a map predicting order of magnitude of heat transfer modification in general configurations.}

Furthermore, we explored different timescales to define the Stokes number and heat mixing parameter, namely large-eddy turnover time and Kolmogorov time.
Our results indicate that macroscopic-averaged heat transfer correction coefficient, $\varphi$, is best described by these parameters \red{(i.e. independent of system Reynolds number)} when they are nondimensionalized based on large eddy time. \red{Investigation of cases with and without momentum two-way coupling between the two phases suggests that the macroscopic heat transfer correction coefficient, $\phi$, is insensitive to modulation of turbulence by the particles.}

A future application of this study is to provide closure to heat transfer terms in \red{subgrid-scale (SGS)} models that do not directly capture particle clustering. For example, Reynolds-averaged Nervier Stokes (RANS) models only represent the ensemble-averaged velocity fields. Therefore, even the most accurate particle solver can provide mean particle number density, while missing the clustering effect. This is a suitable situation for application of the proposed model, by which the heat transfer terms can be closed using the available mean particle number density and mean turbulence dissipation rate (provided by the turbulence model). Future research can investigate wether the presented approach can be adopted in the context of large-eddy simulations (LES), where the input parameters are defined based on sub-grid turbulence dissipation rate.

This work can be extended by relaxing some of our assumptions such as considering temperature dependent thermodynamical properties, inhomogeneous flows, and compressibility effects in case of extreme heating. Furthermore, dependence of our model parameter $\varphi$ on the Reynolds number can be further studied by considering flows with larger values of Reynolds number. 

We would like to acknowledge the support by the US Department of Energy under the Predictive Science Academic Alliance Program 2 (PSAAP2) at Stanford University. 

%

\end{document}